\documentclass[11pt]{article}
\usepackage{graphicx}
\usepackage{amsmath}
\usepackage{xspace}
\usepackage{ptdr-definitions}

\newcommand{\ST}{\ensuremath{S_\mathrm{T}}\xspace}

\providecommand{\tauh}{\ensuremath{\tau_\mathrm{h}}\xspace}

\title{Search for R-parity violating Supersymmetry using the CMS detector}
\author{
\large
\textsc{Fedor Ratnikov for the CMS collaboration}\thanks{talk presented at the LHCP 2013 Conference in Barcelona, Spain, May 13-18th, 2013}\\[2mm] 
\normalsize Karlsrihe Institute of Technology, Karlsruhe, Germany \\ 
\normalsize Institute of Theoretical and Experimental Physics, Moscow, Russia \\ 
\normalsize Fedor.Ratnikov@cern.ch 
\vspace{-5mm}
}
\date{}

\begin{document}
\maketitle
\begin{abstract}
  In this talk, the latest results from CMS on R-parity violating Supersymmetry are reviewed. We present results using up to 20/fb of data from the 8 TeV LHC run of 2012. Interpretations of the experimental results in terms of production of squarks, gluinos, charginos, neutralinos, and sleptons within RP violating susy models are presented. 
\end{abstract}

\section{Introduction}
\label{intro}
Supersymmetry (SUSY) \cite{Nilles_1983ge, Haber_1984rc} is an attractive extension of the Standard Model. 
It provides natural 
coupling unification, dynamic electroweak symmetry breaking and a solution to the hierarchy problem.
R-parity is assigned to fields as $R_p=(-1)^{3B+L+s}$ where $B$, $L$, and $s$ are baryon and lepton numbers,
and spin of the particle respectively. In models with conserved R-parity superpartners may only be 
produced in pairs, and the lightest superpartner (LSP) is stable. 
However R-parity conservation is not a universal property of SUSY models.
The most general gauge-invariant and renormalizable
superpotential consists of the R-parity conserving (RPC) main part, and may also contain extra R-parity 
violating (RPV) terms \cite{Martin_1997ns}:
\begin{eqnarray}
\label{eq:lagrangian}
W_{\Delta L = 1} & = & \frac{1}{2} \lambda_{ijk}L_iL_j\bar{e}_k +  \lambda'_{ijk}L_iQ_j\bar{d}_k + \mu'_iL_iH_u \\
W_{\Delta B = 1} & = & \frac{1}{2} \lambda''_{ijk}u_id_j\bar{d}_k
\end{eqnarray}

The presence of non-vanishing RPV terms leads to the the LSP becoming unstable, decaying to 
standard model (SM) particles. Therefore many SUSY analyses, which are based on the expectation of high missing 
transverse energy in SUSY events from non-observed stable LSPs, 
are not sensitive to RPV SUSY models.

Recent CMS analyses \cite{CMS-PAS-SUS-13-003, CMS-PAS-SUS-13-005, CMS-PAS-SUS-13-010} 
are focused on studying the lepton number violating terms
 $\lambda_{ijk}L_iL_j\bar{e}_k$ and $\lambda'_{ijk}L_iQ_j\bar{d}_k$, which cause specific signatures involving
leptons in events produced in pp collisions at LHC.
Section \ref{sec:SUS-13-005} discusses the search for resonant production and the following decay of $\tilde{\mu}$ which
is caused by $\lambda'_{211} \neq 0$. Section \ref{sec:SUS-13-003} addresses a search for multi-lepton signatures caused
by LSP decays due to various $\lambda$ and $\lambda'$ terms. Finally in Section \ref{sec:SUS-13-010} we discuss
the possibility of the generic model independent search for RPV SUSY in 4-lepton events.

\section{Detector, trigger, and object selection}
\label{detector}

The central feature of the CMS apparatus is a superconducting
solenoid, 6~m in internal diameter, providing a magnetic field of
3.8~T. Within the field volume there are a silicon pixel and strip
tracker, a crystal electromagnetic calorimeter, and a
brass-scintillator hadron calorimeter. Muons are measured in
gas-ionization detectors embedded in the steel return yoke. Extensive
forward calorimetry complements the coverage provided by the barrel
and endcap detectors. A more detailed description of the CMS detector can be found in
Ref.~\cite{Chatrchyan_2008aa}.

Events from pp interactions must satisfy the requirements of a
two-level trigger system. The first level performs a fast selection
for physics objects (jets, muons, electrons, and photons) above
certain thresholds.  The second level performs a full event
reconstruction.
The principal trigger used for these analyses requires presence of at least
two light leptons, electrons or muons. Detailed trigger conditions
and off-line event selections are described in the corresponding 
Ref.~\cite{CMS-PAS-SUS-13-003, CMS-PAS-SUS-13-005, CMS-PAS-SUS-13-010}.
 
\section{Search for resonant second generation slepton production}
\label{sec:SUS-13-005}

\begin{figure}[htbp]
\centering
\includegraphics[width=0.49\textwidth,clip]{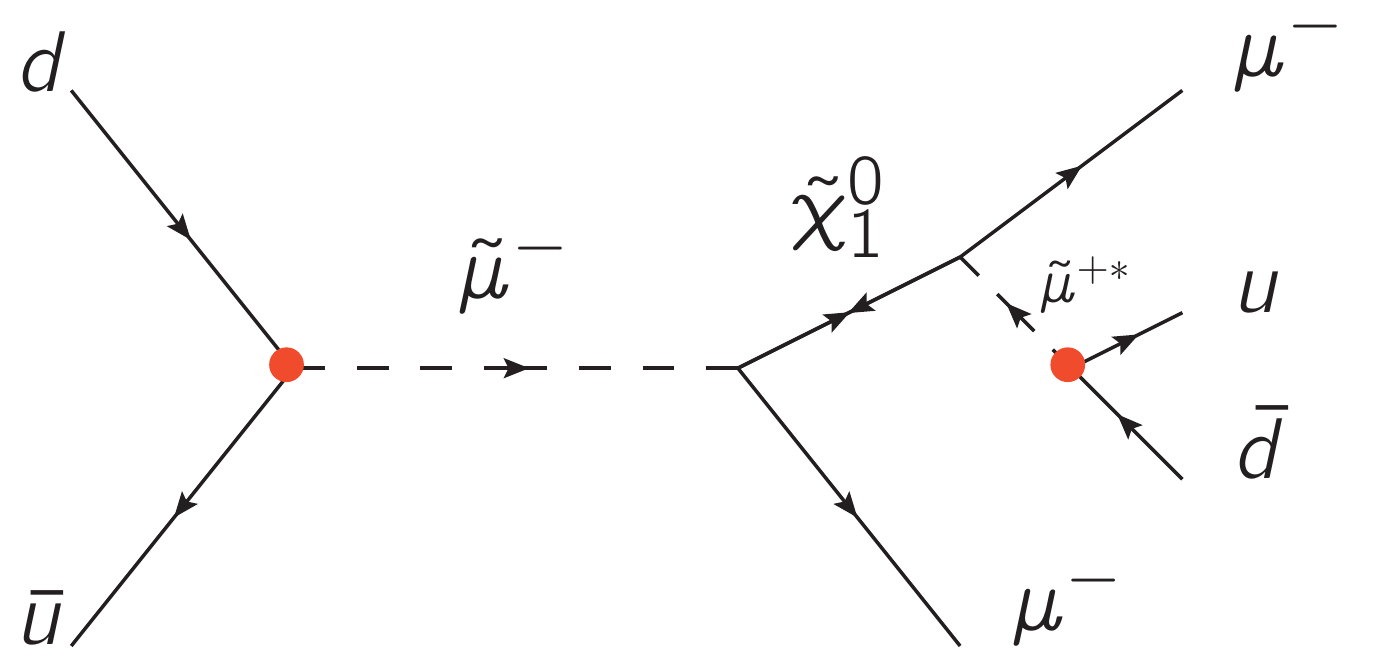}
\hfill
\includegraphics[width=0.49\textwidth,clip]{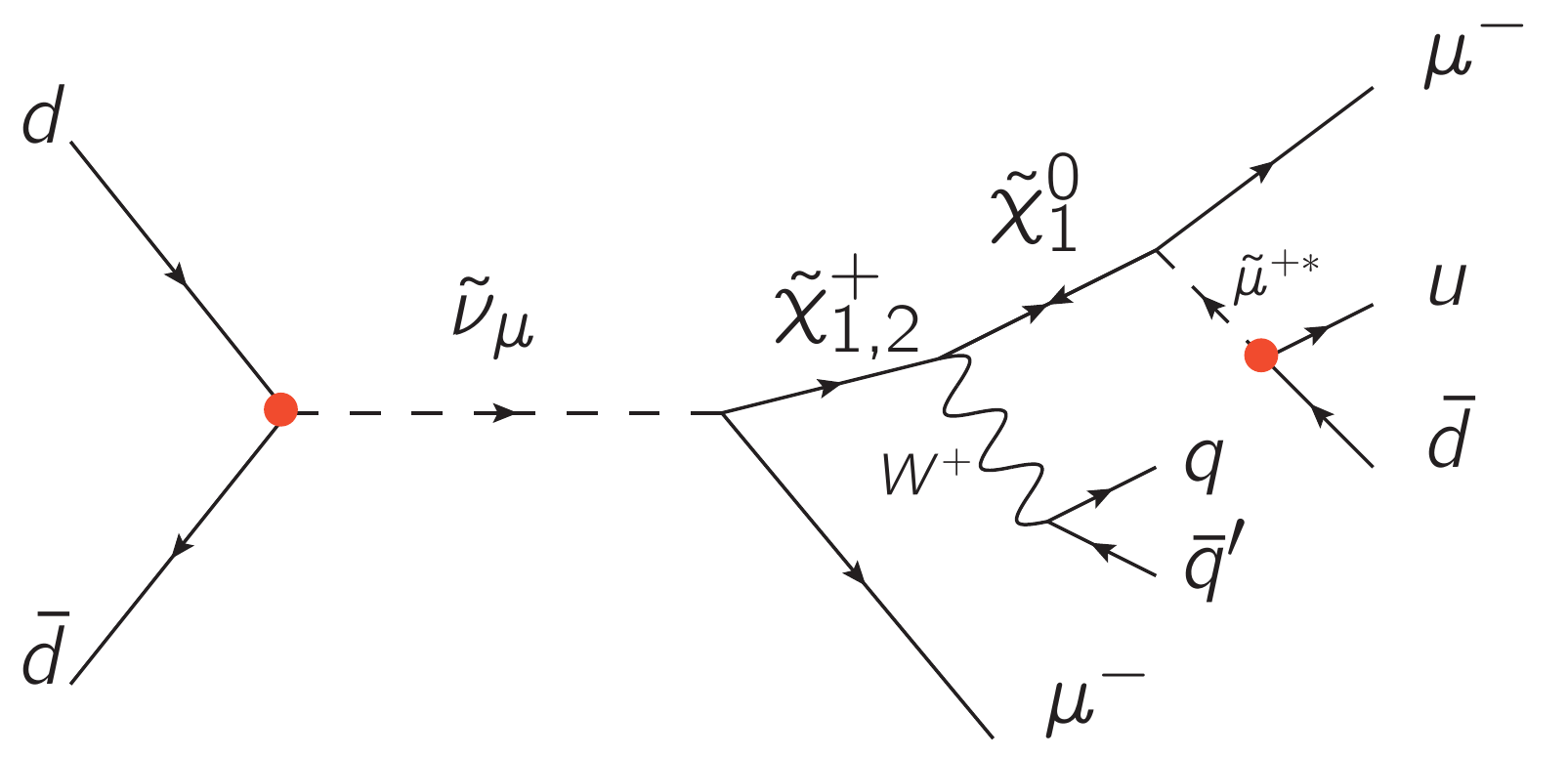}
\caption{Resonant smuon (left) and sneutrino (right) production and typical decay chain into a final 
state with two same-sign muons and two jets. The R-parity violating vertices are marked by a red dot.}
\label{fig:1}       
\end{figure}

\begin{figure}[htbp]
\centering
\includegraphics[width=0.49\textwidth,clip]{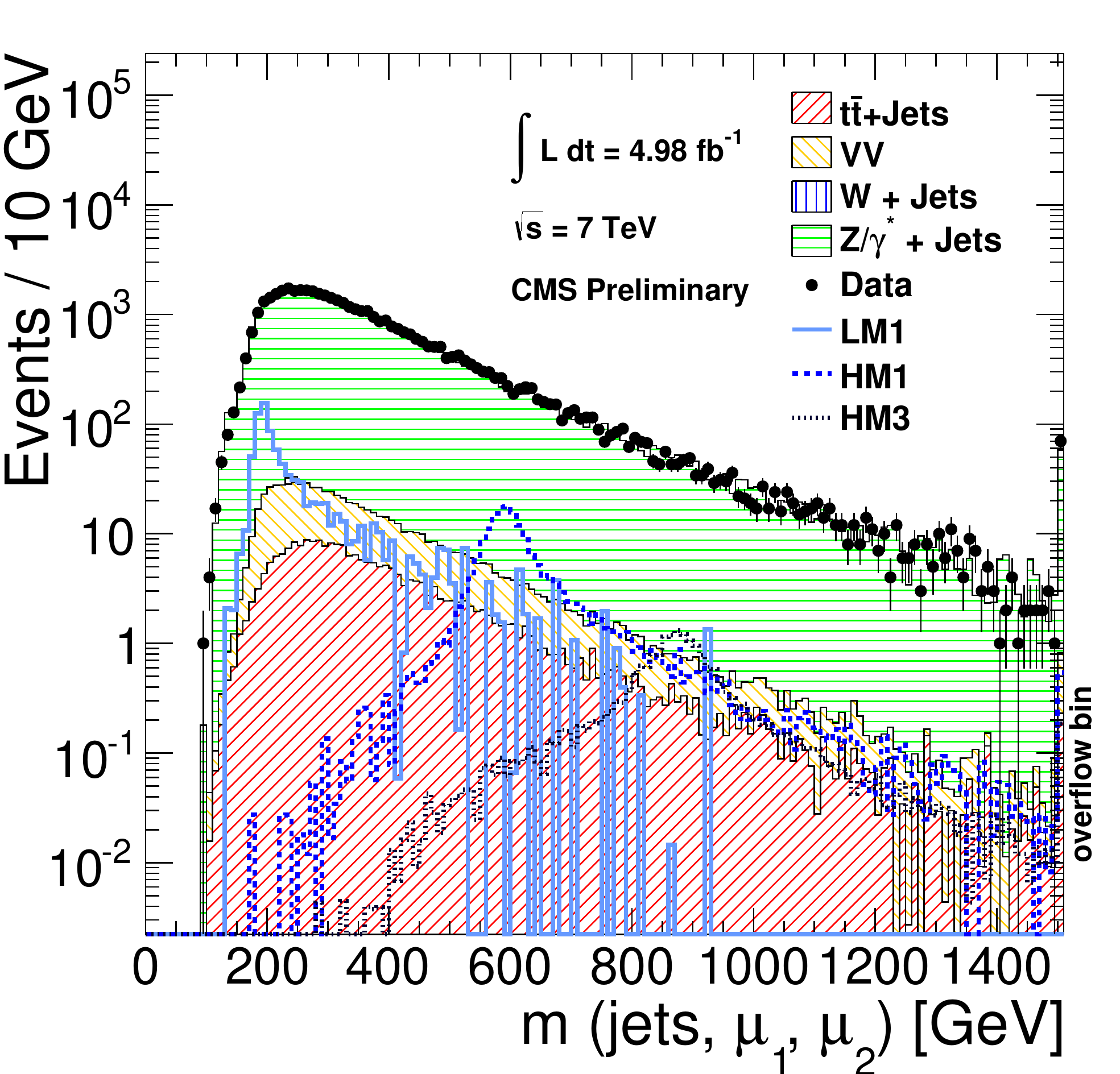}
\hfill
\includegraphics[width=0.49\textwidth,clip]{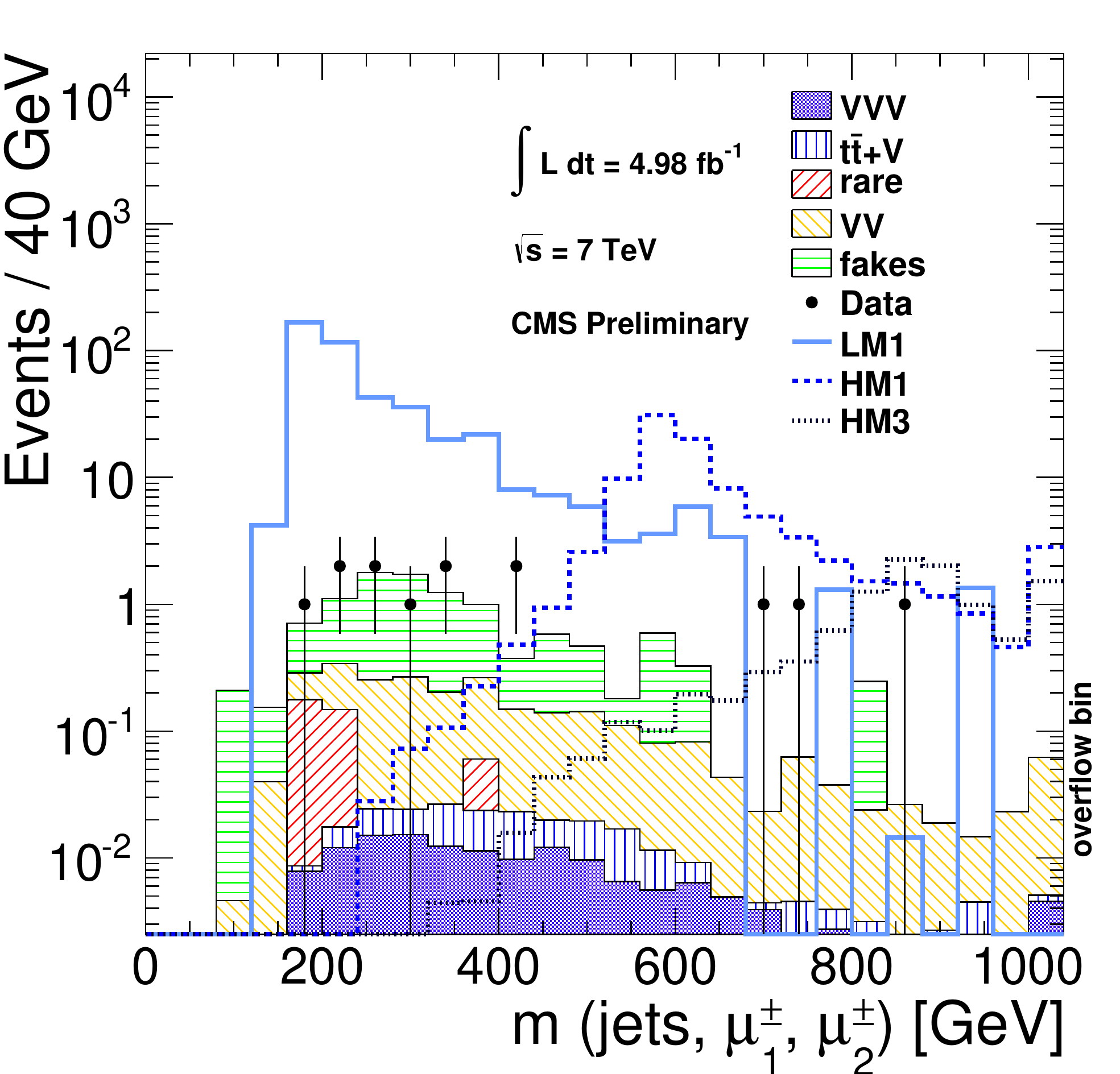}
\caption{Invariant mass of the two muons and jets before (left) and after (right) applying the b-tag veto 
and same-sign muon requirement. Data are compared to the expectation from the simulation (left) 
and measured backgrounds (right). Signal distributions are shown for three different kinematic configurations 
for a coupling value of $\lambda'_{211}=0.01$. 
}
\label{fig:2}       
\end{figure}

\begin{table}[htbp]
\caption{Event yields with systematic uncertainties after selection requirements,
  broken down in individual Standard Model background contributions, with
  observed 95\% C.L.\ limits on the number of signal events $N_{sig}$ in total
  and for each signal region.}
\label{tab:events05}
\begin{center}
\begin{small}
\begin{tabular}{l|r|rrr}
process & totals & SR1 & SR2 & SR3 \\
\hline
VVV & 0.15 $\pm$ 0.08 & 0.043 $\pm$ 0.022 & 0.054 $\pm$ 0.028 & $<$0.001 \\
tt+V & 0.11 $\pm$ 0.06 & 0.019 $\pm$ 0.010 & 0.038 $\pm$ 0.020 & 0 \\
rare & 0.36 $\pm$ 0.26 & 0.32 $\pm$ 0.24 & 0.042 $\pm$ 0.042 & $<$0.001 \\
VV & 2.1 $\pm$   1.1 & 0.69 $\pm$ 0.35 & 0.68 $\pm$ 0.34 & 0.003 $\pm$ 0.002 \\
fakes & 8.2 $\pm$   3.0 & 3.5  $\pm$ 1.6 & 1.9  $\pm$ 1.0 & $<$0.001 \\
\hline
$\sum$ & 10.9 $\pm$  3.4 & 4.6 $\pm$  1.6 & 2.7  $\pm$  1.1 & 0.003 $\pm$ 0.002 \\
\hline
data & 13 & 5 & 5 & 0 \\
\hline
95\% C.L. limit on $N_{sig}$ & 11.3 & 6.9 & 8.0 & 2.8 \\
\hline
\hline
process &  & SR4 & SR5 & SR6 \\
\hline
VVV &  & 0.036 $\pm$ 0.018 & 0.010 $\pm$ 0.005 & 0.007 $\pm$ 0.004 \\
tt+V &  & 0.044 $\pm$ 0.023 & 0.006 $\pm$ 0.004 & 0.006 $\pm$ 0.004 \\
rare &  & $<$0.001 & $<$0.001 & $<$0.001 \\
VV &  & 0.49 $\pm$ 0.25 & 0.15 $\pm$ 0.08 & 0.093 $\pm$ 0.050 \\
fakes &  & 2.5  $\pm$ 1.2 & 0.22 $\pm$ 0.23 & $<$0.001 \\
\hline
$\sum$ &  & 3.1 $\pm$  1.2 & 0.39 $\pm$ 0.25 & 0.11 $\pm$ 0.05 \\
\hline
data &  & 0 & 2 & 1 \\
\hline
95\% C.L. limit on $N_{sig}$ &  & 2.9 & 6.0 & 4.6 \\
\end{tabular}
\end{small}
\end{center}
\end{table}

\begin{figure}
\centering
\includegraphics[width=0.70\textwidth,clip]{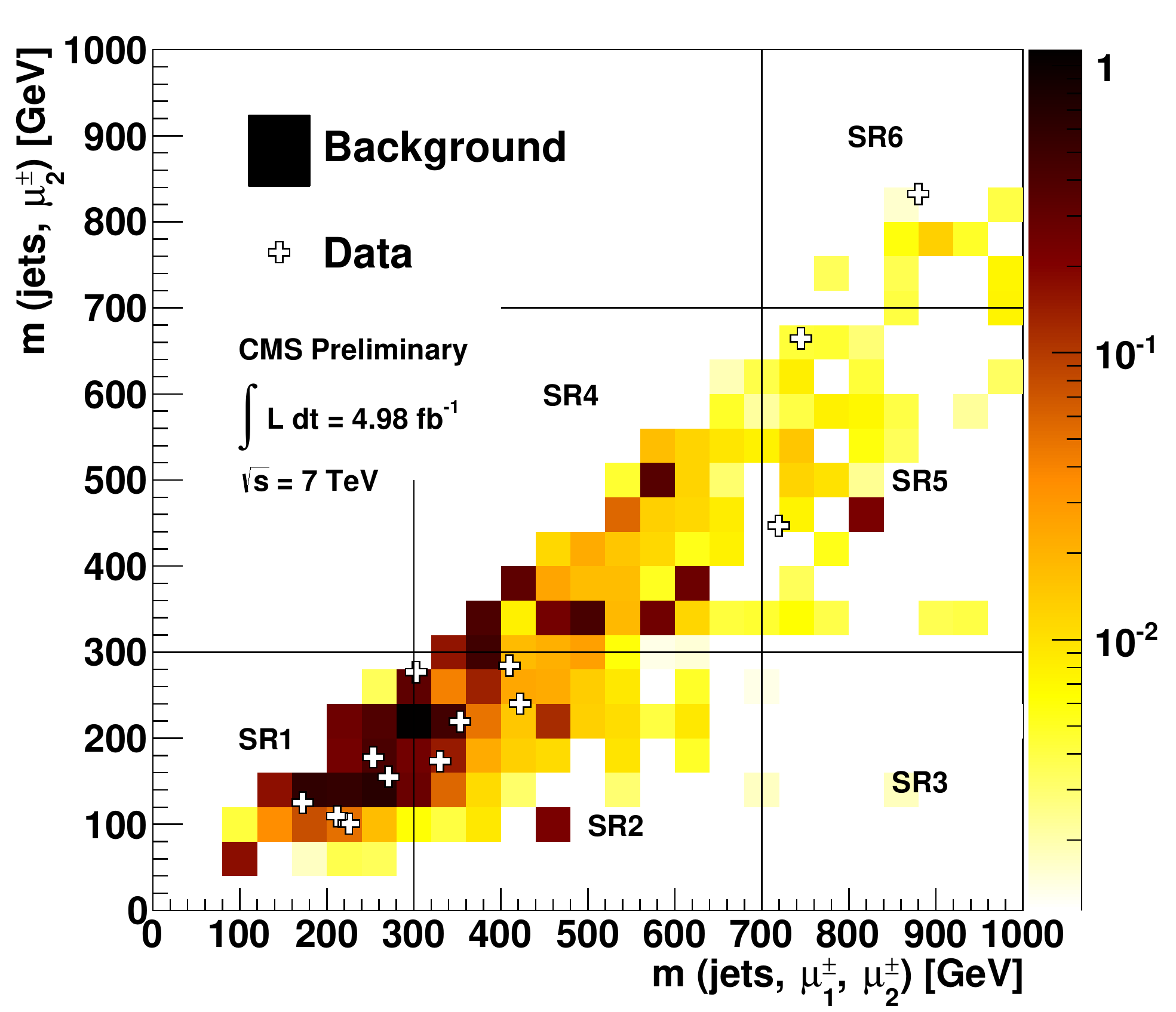}
\caption{Distribution of $m_{\tilde\mu} = m(\text{jets}, \mu_1^\pm, \mu_2^\pm)$
vs. $m_\chi = m(\text{jets}, \mu_2^\pm)$ for the events selected in data
compared to the total background contribution. The crosses represent the data
points and the coloured squares show the expectation from Standard Model
backgrounds.
}
\label{fig:3}       
\end{figure}

\begin{figure}[htbp]
\centering
\includegraphics[width=.44\textwidth,height=0.30\textwidth,clip]{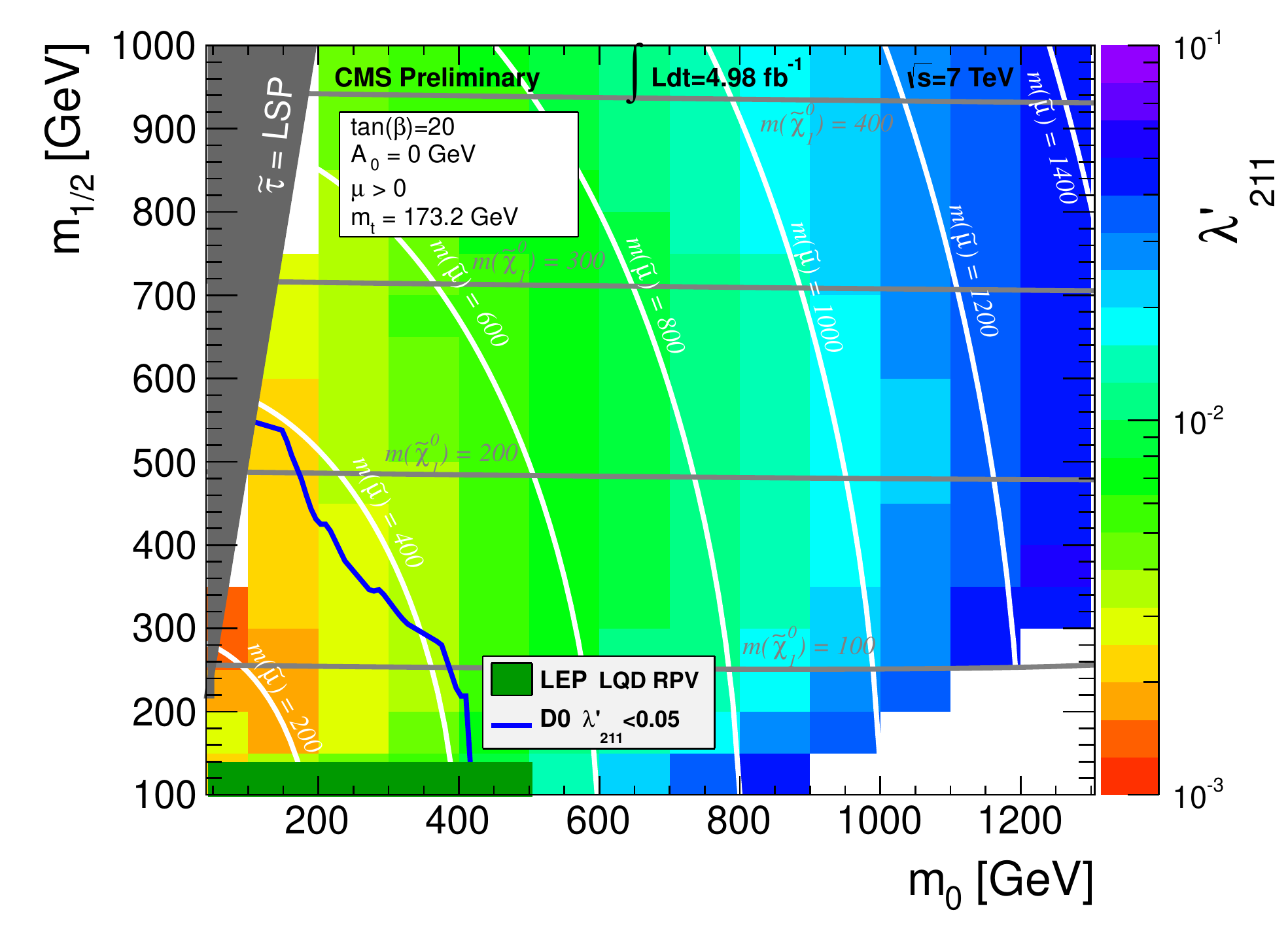}
\hfill
\includegraphics[width=.44\textwidth,height=0.32\textwidth,clip]{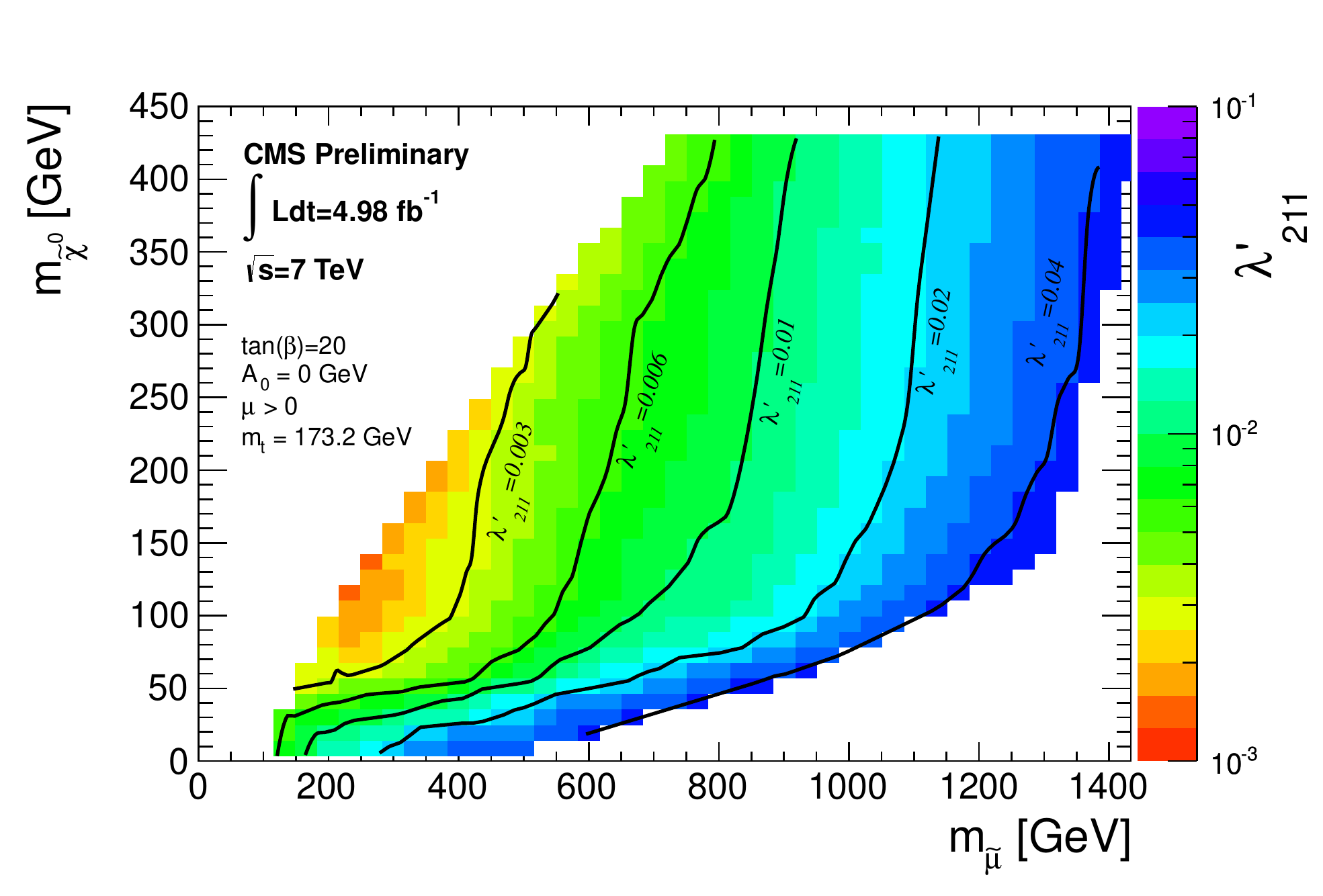}
\caption{Left: observed 95\% CL upper limits on
  $\lambda'_{211}$ as a function of $m_0$ and $m_{1/2}$ for $A_0 = 0$,
  sign~$(\mu) = +1$ and $\tan\beta = 20$. Right: mSUGRA
  limits expressed in the parameter space of the neutralino mass
  $m_{\tilde{\chi}^0_1}$ and smuon mass $m_{\tilde{\mu}}$.
}
\label{fig:4}       
\end{figure}

This search which is described in details in Ref.~\cite{CMS-PAS-SUS-13-005}, 
extends the results from a previous search by the D\O\
collaboration~\cite{Abazov_2006ii} and is complementary to searches for RPV SUSY 
performed by
the LEP experiments~\cite{Barbier_2004ez}. The search concentrates on final
states with two muons and at least two jets. Fig.~\ref{fig:1}
illustrates the simplest possible Feynman diagrams leading to this final
state, which is experimentally interesting because the presence of two muons
allows to discriminate the signal from background processes. One of the muons
is expected to be produced by the resonant slepton while the other muon and
two quarks resulting in jets are expected to be produced in the subsequent
decay of the neutralino LSP. Due to the Majorana nature of the LSP, the two
muons have the same charge with about 50\% probability, which allows to
discriminate further against the background. Due to the larger valence
$\mathrm{u}$-quark content of the initial state protons the configuration with
two positively charged muons is about twice as likely as the configuration with
two negatively charged muons. The kinematics of this signal is characterized
by no missing transverse energy within the detector resolution. 
For the purpose of this analysis we select events with two same-sign isolated 
muons with $\ensuremath{p_{\mathrm{T}}}\xspace >20$ and $\pt>15$\GeV for the first and second muon respectively. 
In addition at least two jets with $\pt>30$\GeV, no $\mathrm{b}$-jets, 
and $\MET<50$\GeV are required.
After this selection, two main background
components remain: low cross section backgrounds containing two
prompt same-sign leptons such as production of multiple bosons, and
backgrounds with high cross-section where leptons from semileptonic decays of
$\mathrm{c}$ or $\mathrm{b}$-hadrons or other charged particles are wrongly
identified as prompt leptons. The first contribution is estimated from the simulation.
The latter contribution is difficult to model in simulation, thus it is 
estimated using data. Fig.~\ref{fig:2} illustrates the expected backgrounds
before and after the requirement of the two same-sign muons and the $\mathrm{b}$-jets veto.

The 13 events observed in Fig \ref{fig:2} (right) are further investigated 
using their 2D distribution in parameters $m_{\tilde\mu} = m(\text{jets}, \mu_1^\pm, \mu_2^\pm)$
vs. $m_\chi = m(\text{jets}, \mu_2^\pm)$, where $\mu_1^\pm$ denoting the muon with higher \pt.
Fig.~\ref{fig:3} overlays the observed events with the expected background contributions,
and describes six exclusive search regions used for the interpretation of this analysis.
Table \ref{tab:events05} presents the observations, expected backgrounds, and respective
upper limits for all search regions. The observations are consistent with the corresponding
background estimations, therefore results are combined to put limit on $\lambda'_{211}$
for different mSugra models in Fig. \ref{fig:4}.  

\begin{table*}[!ht]
\begin{center}
\caption{Observed yields for three- and four- lepton events from 19.5 fb$^-1$ recorded in 2012. The channels are split by the total number of leptons (N$_\mathrm{L}$), the number of $\tauh$ candidates (N$_{\tau}$), and the $\ST$. Expected yields are the sum of simulation and estimates of backgrounds from data in each channel. SR1--SR4 require a $\cPqb$-tagged jet and veto events containing \cPZ~bosons. SR5--SR8 contain events that either contain a \cPZ~boson or have no $\cPqb$-tagged jet. The channels are mutually exclusive. The uncertainties include statistical and systematic uncertainties. The $\ST$ values are given in\GeV.
\label{tab:results03}}
\begin{tiny}
\begin{tabular}{ccccccccccccc}
SR & N$_\mathrm{L}$ & N$_{\tau}$ & \multicolumn{2}{c}{$0 < \ST < 300$} &\multicolumn{2}{c}{$300 < \ST < 600$}& \multicolumn{2}{c}{$600 < \ST < 1000$}& \multicolumn{2}{c}{$1000 < \ST < 1500$} & \multicolumn{2}{c}{$\ST > 1500$} \\
\hline
& & & obs & exp & obs & exp & obs & exp & obs & exp & obs & exp\\
\hline
SR1 & 3 & 0 & 116 & 123 $\pm$ 50 & 130 & 127 $\pm$ 54 & 13 & 18.9 $\pm$ 6.7 & 1 & 1.43 $\pm$ 0.51 & 0 & 0.208 $\pm$ 0.096 \\
SR2 & 3 & $\ge1$ & 710 & 698 $\pm$ 287 & 746 & 837 $\pm$ 423 & 83 & 97 $\pm$ 48 & 3 & 6.9 $\pm$ 3.9 & 0 & 0.73 $\pm$ 0.49 \\
SR3 & 4 & 0 & 0 & 0.186 $\pm$ 0.074 & 1 & 0.43 $\pm$ 0.22 & 0 & 0.19 $\pm$ 0.12 & 0 & 0.037 $\pm$ 0.039 & 0 & 0.000 $\pm$ 0.021 \\
SR4 & 4 & $\ge1$ & 1 & 0.89 $\pm$ 0.42 & 0 & 1.31 $\pm$ 0.48 & 0 & 0.39 $\pm$ 0.19 & 0 & 0.019 $\pm$ 0.026 & 0 & 0.000 $\pm$ 0.021 \\
SR5 & 3 & 0 &--- &--- &--- & ---& 165 & 174 $\pm$ 53 & 16 & 21.4 $\pm$ 8.4 & 5 & 2.18 $\pm$ 0.99 \\
SR6 & 3 & $\ge1$  &--- &--- & ---&---& 276 & 249 $\pm$ 80 & 17 & 19.9 $\pm$ 6.8 & 0 & 1.84 $\pm$ 0.83 \\
SR7 & 4 & 0 &--- &--- &--- &--- & 5 & 8.2 $\pm$ 2.6 & 2 & 0.96 $\pm$ 0.37 & 0 & 0.113 $\pm$ 0.056 \\
SR8 & 4 & $\ge1$  &--- &--- &--- &--- & 2 & 3.8 $\pm$ 1.3 & 0 & 0.34 $\pm$ 0.16 & 0 & 0.040 $\pm$ 0.033 \\
\end{tabular}
\end{tiny}
\end{center}
\end{table*}

\section {Search for R-parity violating SUSY in multileptons with $\cPqb$-tagged jets}
\label{sec:SUS-13-003}

\begin{table}[htbp]
\caption{\label{tab:stopdecays}
Kinematically allowed stop decay modes with RPV coupling $\Lamp_{233}$.
The allowed neutralino decay modes for $ m_{\cPqt} < m_{\PSGczDo} < m_{\stone}$
are $\PSGczDo \to \mu \cPqt \cPaqb$ and $\nu \cPqb \cPaqb$.
}
\centering
\begin{tabular}{ccc}
Label & Kinematic region & Decay mode\\
\hline
A & $ m_{\cPqt} < m_{\stone} < 2m_{\cPqt} ,  m_{\PSGczDo} $ & $\stone \rightarrow  \cPqt \nu \cPqb \cPaqb $ \\
B & $ 2 m_{\cPqt} < m_{\stone} <  m_{\PSGczDo} $ & $\stone \rightarrow \cPqt \mu \cPqt \cPaqb  $ or $ \cPqt \nu \cPqb \cPaqb $ \\
C & $ m_{\PSGczDo} < m_{\stone} < m_{\Wpm} + m_{\PSGczDo} $ & $\stone \rightarrow \ell \nu \cPqb \PSGczDo $ or $ \cmsSymbolFace{jj}\cPqb \PSGczDo $ \\
D & $m_{\Wpm} + m_{\PSGczDo} < m_{\stone} < m_{\cPqt} + m_{\PSGczDo} $ & $\stone \rightarrow \cPqb \Wpm \PSGczDo$ \\
E & $m_{\cPqt} + m_{\PSGczDo} < m_{\stone} $ & $\stone \rightarrow \cPqt \PSGczDo$ \\
\end{tabular}
\end{table}

\begin{figure}[htbp]
   \caption{The 95\% confidence level limits in the stop and bino mass plane for models with RPV couplings $\Lam_{122}$, $\Lam_{233}$, and $\Lamp_{233} $. For the couplings $\Lam_{122}$ and $\Lam_{233}$, the region to the left of the curve is excluded. For $\Lamp_{233}$, the region inside the curve is excluded.
The different regions, A, B, C, D, and E, for the $\Lamp_{233}$ exclusion result from different stop decay products as explained in Table~\ref{tab:stopdecays}.
      }
\centering
\includegraphics[width=0.49\textwidth]{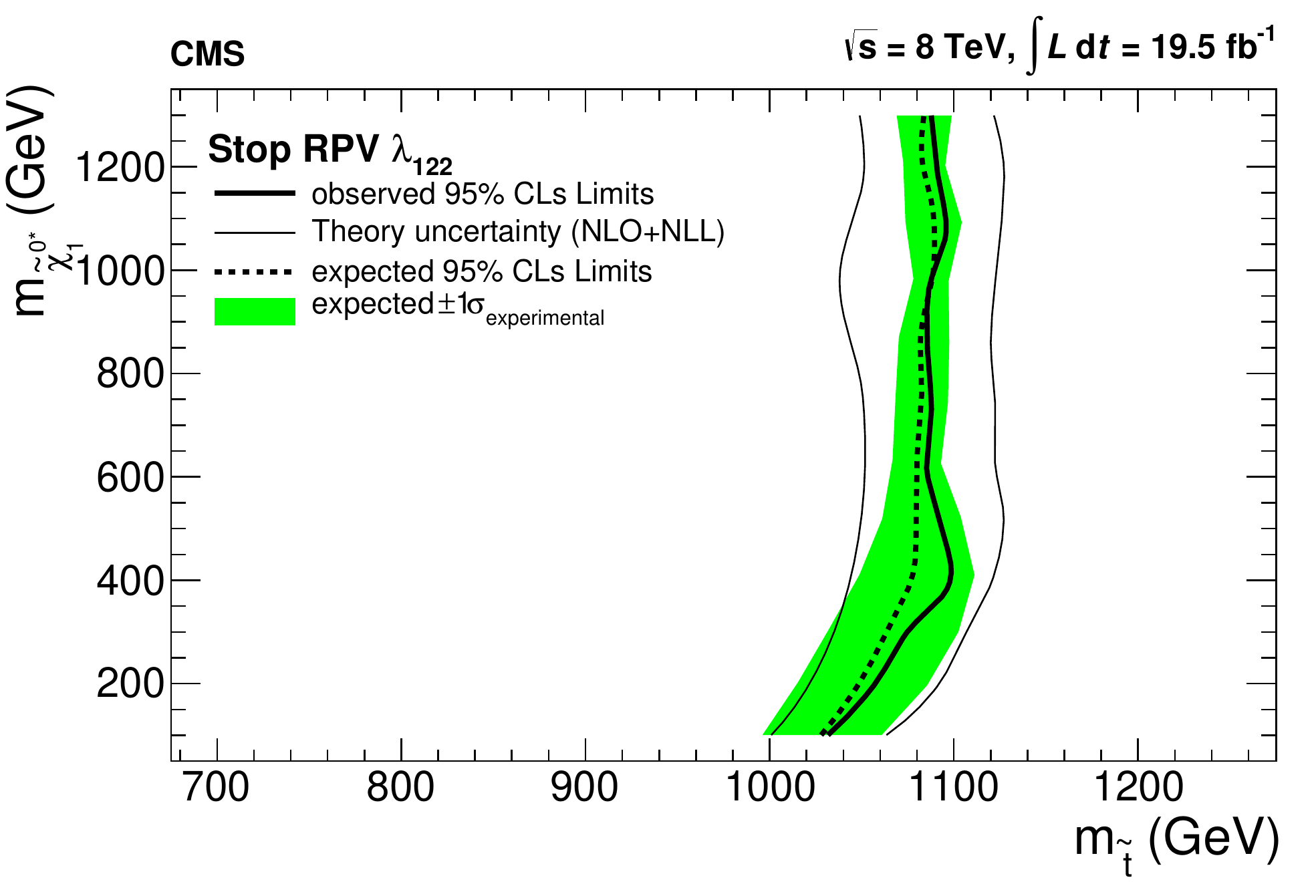}
\includegraphics[width=0.49\textwidth]{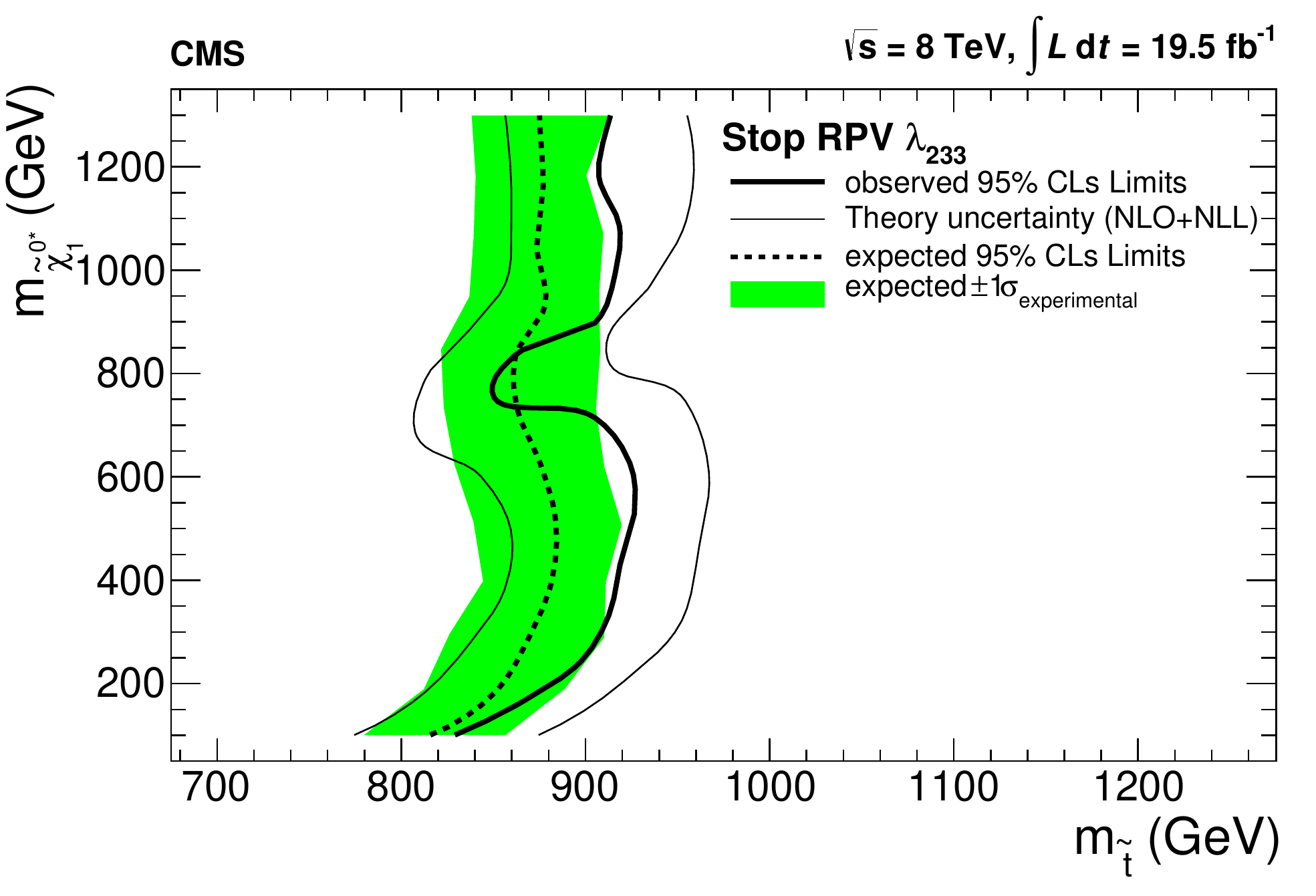}
\includegraphics[width=0.49\textwidth]{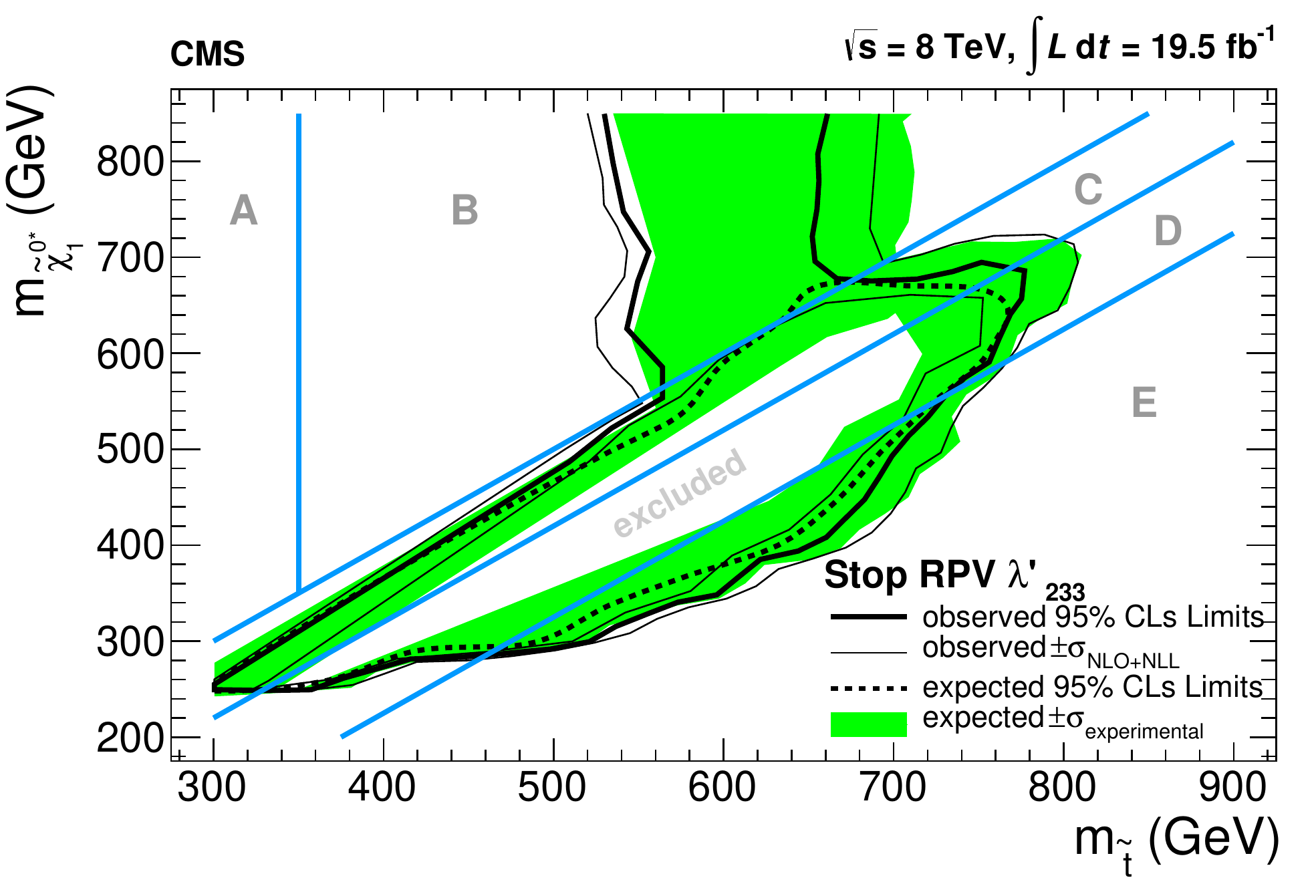}
   \label{fig:stopRPV}
\end{figure}

Among modern SUSY models, ``natural" supersymmetry refers to those characterized by a relatively 
small fine tuning to describe particle spectra. It requires top squarks (stops), 
to be lighter than about 1\TeV. The introduction of RPV does not preclude a natural 
hierarchy and allows the constraints on the stop mass to be relaxed \cite{Evans_2012bf}.

The analysis \cite{CMS-PAS-SUS-13-003} searches for pair production of top squarks with RPV decays of the lightest sparticle, 
using multilepton events and $\cPqb$-tagged jets. 
It addresses terms $\lambda_{ijk}$ and $\lambda'_{ijk}$ in Eqn.~\ref{eq:lagrangian}. 

We select events with three or more leptons (including tau leptons) that are accepted by a trigger required two light leptons, which may be electrons or muons. At least one electron or muon in each event is required to have transverse momentum of 
$\pt > 20$\GeV. Additional electrons and muons must have $p_{\rm T} > 10$\GeV.
The majority of hadronic decays of tau leptons ($\tauh$) yield either a single charged track (one-prong) or three 
charged tracks (three-prong), occasionally with additional electromagnetic energy from neutral pion decays. 
We use one- and three-prong $\tauh$ candidates that have $\pt > 20$\GeV. 
Leptonically decaying taus are included with other electrons and muons.
The \MET is not a good discriminator for RPV SUSY search. Instead we use the \ST variable, which is the scalar sum of
\MET and the transverse energy of jets with $\pt>30$\GeV and charged leptons, to provide separation 
between signal and the Standard Model backgrounds.

Irreducible Standard Model backgrounds are estimated from simulation. Contributions from fakes for electrons, muons
and taus are obtained using data-driven methods.

Observed events are classified into eight topologies according to the number of observed light leptons and
the presence of hadronic tau in event. Every topology is further split into five search regions
according to the \ST value. Table~\ref{tab:results03} summarizes observations and expected contributions
for different search regions used in this analysis.

We generate simulated samples to evaluate models with simplified mass spectra and the only non-zero leptonic RPV couplings 
$\Lam_{122}$ or $\Lam_{233}$. The stop masses in these samples range from 700--1250\GeV in 50\GeV steps, 
and bino masses range from 100--1300\GeV in 100\GeV steps. In a model with only the semi-leptonic RPV coupling 
$\Lamp_{233}$, we use stop masses 300--1000\GeV in 50\GeV steps and bino masses  200--850\GeV in 50\GeV steps. 
In both cases, slepton and sneutrino masses are 200\GeV above the bino mass. 
Other particles are irrelevant to the results for these models.

No significant excess is observed in data. The observations from Table~\ref{tab:results03} are combined into 
exclusions for the corresponding models in Fig.~\ref{fig:stopRPV}.

\section {Generalization of Unstable LSP Search}
\label{sec:SUS-13-010}


The analysis described in details in Ref.~\cite{CMS-PAS-SUS-13-010} presents a new approach to a generic interpretation
of experimental results.
The focus of this analysis is the lepton number violating term $\lambda_{ijk}L_iL_j\bar{e}_k$, which
causes the LSP in such a ``Leptonic-RPV'' (LRPV) SUSY model to decay into leptons. 
SUSY particles are produced in pairs, thus a non-zero $\lambda$-term would lead 
to events with 4 charged leptons produced in LSP decays. 
Recent searches at the Tevatron \cite{Abazov_2006nw} and 
LHC \cite{ATLAS_2012kr, CMS-PAS-SUS-13-003} placed limits on 
$\lambda$. The main challenge of RPV SUSY searches is that the RPV term exists on top of some underlying RPC SUSY
model, with properties which are currently barely constrained. 
The analyses mentioned above resolve this problem
by exploring RPV on top of very specific RPC SUSY models.
In this analysis we pursue a significantly less model dependent approach.
We require the presence of 4 isolated leptons in the event, as a direct signature of the LRPV SUSY.
No other restriction is applied, so the selection efficiency is not directly affected by the underling
SUSY event. Irreducible Standard Model backgrounds are estimated from simulation, estimations of fakes
are data-driven. The main background for 4-lepton events is found to be ZZ production, so for every
event the variable $M_1$ is calculated as the invariant mass of same-flavor opposite-sign lepton pair that is
closest to the mass of Z-boson. $M_2$ is then calculated as the invariant mass of the remaining lepton pair.

\begin{table}[!htdp]
  \caption{Observed events and expected background contributions. $M_1$ and $M_2$ intervals are in \GeV.}
  \label{tab:observations10}
  \begin{center}
    \begin{tabular}{|c|l|c|c|c|}\hline
    \multicolumn{2}{|c|}{} & $M_1<76$ & $76<M_1<106$ & $M_1>106$ \\ \hline \hline 
                  & all backgrounds & 1.4$\pm$0.5 & 18$\pm$4 & 0.47$\pm$0.10 \\ \cline{2-5} 
    $M_2>106$ & observed        & 0           & 20         & 0  \\ \hline \hline                    
                  & all backgrounds & 0.52$\pm$0.30& 153$^*$   & 0.16$\pm$0.06 \\ \cline{2-5} 
$76<M_2<106$  & observed        & 0           & 160        & 0 \\ \hline \hline                    
                  & all backgrounds & 10.4$\pm$2.0 & 35$\pm$8  & 1.0$\pm$0.2 \\ \cline{2-5} 
   $M_2<76$ & observed         & 14  & 30 & 1 \\ \hline \hline                    
    \end{tabular}
  \end{center}
  $^*$ ZZ prediction in ``in~Z'':``in~Z'' region is based on MC normalized to CMS ZZ production cross section measurement, which 
   is correlated with observation in ``in~Z'':``in~Z'' region of this analysis.
\end{table}

Table \ref{tab:observations10} presents the observations and expected backgrounds in different regions
in $M_1:M_2$ space. Observations and expectations are consistent in all regions.
Based on the occupancy of different regions for typical ZZ production 
events, the signal region is defined as ``$M_1$ above Z'' or
 ``$M_1$ below Z and $M_2$ above Z''. Then the upper limit on cross section times integrated luminosity
times efficiency ($\sigma\times\mathcal{L}\times\varepsilon$) 
for any physics process beyond the SM contributing to this search region is 3.4 events. The expected
upper limit for this observation is 4.7 events.
The leptonic decay of the pair of LRPV neutralinos leads to 4 prompt leptons. The kinematics of these leptons
are in general driven by the momentum distribution of the decaying neutralinos and their mass. 
In most scenarios the lepton momentum is well above threshold, which results in high 
efficiency. However the following effects could reduce the total efficiency:
\begin{itemize}
\item the presence of other leptons in the event, which affects the efficiency through 
the 4-lepton requirement, as well as the calculation of the $M_1$ and $M_2$ quantities;
\item the electron and/or muon objects reconstruction efficiency which is dependent on $\eta$ and $p_T$;
\item the isolation efficiency, which is correlated with the occupancies around the observed prompt leptons.
\end{itemize}
The presence of an extra lepton in the SUSY event, in addition to the 4 leptons produced from neutralino
decays, could veto the event. We observe no events containing 5 isolated leptons. 
Thus, the potential presence of additional leptons in fact does not significantly affect the measurement.

\begin{figure}[!htdp]
  \begin{center}
    \includegraphics[width=0.5\textwidth]{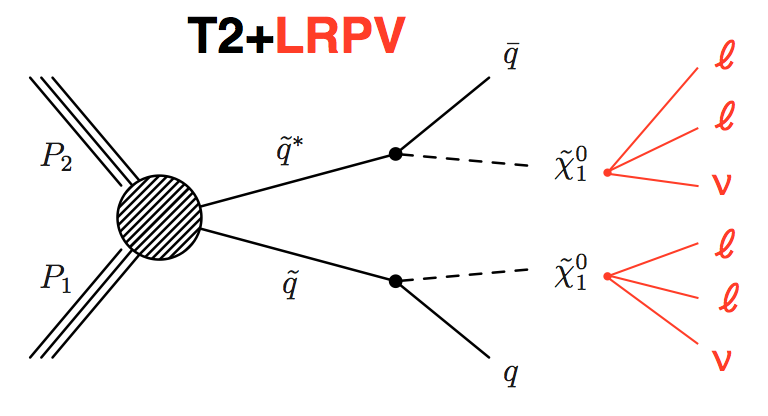}
    \caption{
      LRPV extensions to Simplified Model \cite{Alves_2011wf}. 
      The T2 RPC simplified model is squark pair production, 
      with $\tilde{q}\rightarrow q \tilde{\chi}_1^0$, and $m(\tilde{g}) \gg m(\tilde{q})$. 
     The neutralinos decay to two charged leptons and a neutrino via an LRPV term.
    }
    \label{fig:T2LRPV}
  \end{center}
\end{figure}

To evaluate the dependency of the lepton reconstruction efficiency and the efficiency of analysis selections from 
details of kinematic distributions of decaying neutralinos,  we consider
two extreme cases of LRPV neutralino production:
\begin{itemize}
\item a simplified model with SUSY particles produced via a squark-anti-squark pair, 
with the neutralino coming from a two-body decay $\tilde{q}\rightarrow q \tilde{\chi}_1^0$,
as presented in Fig.~\ref{fig:T2LRPV};
\item a pair of neutralinos produced in rest in the center of the CMS detector.
\end{itemize}

\begin{figure}[!htdp]
  \begin{center}
    \includegraphics[width=0.49\textwidth]{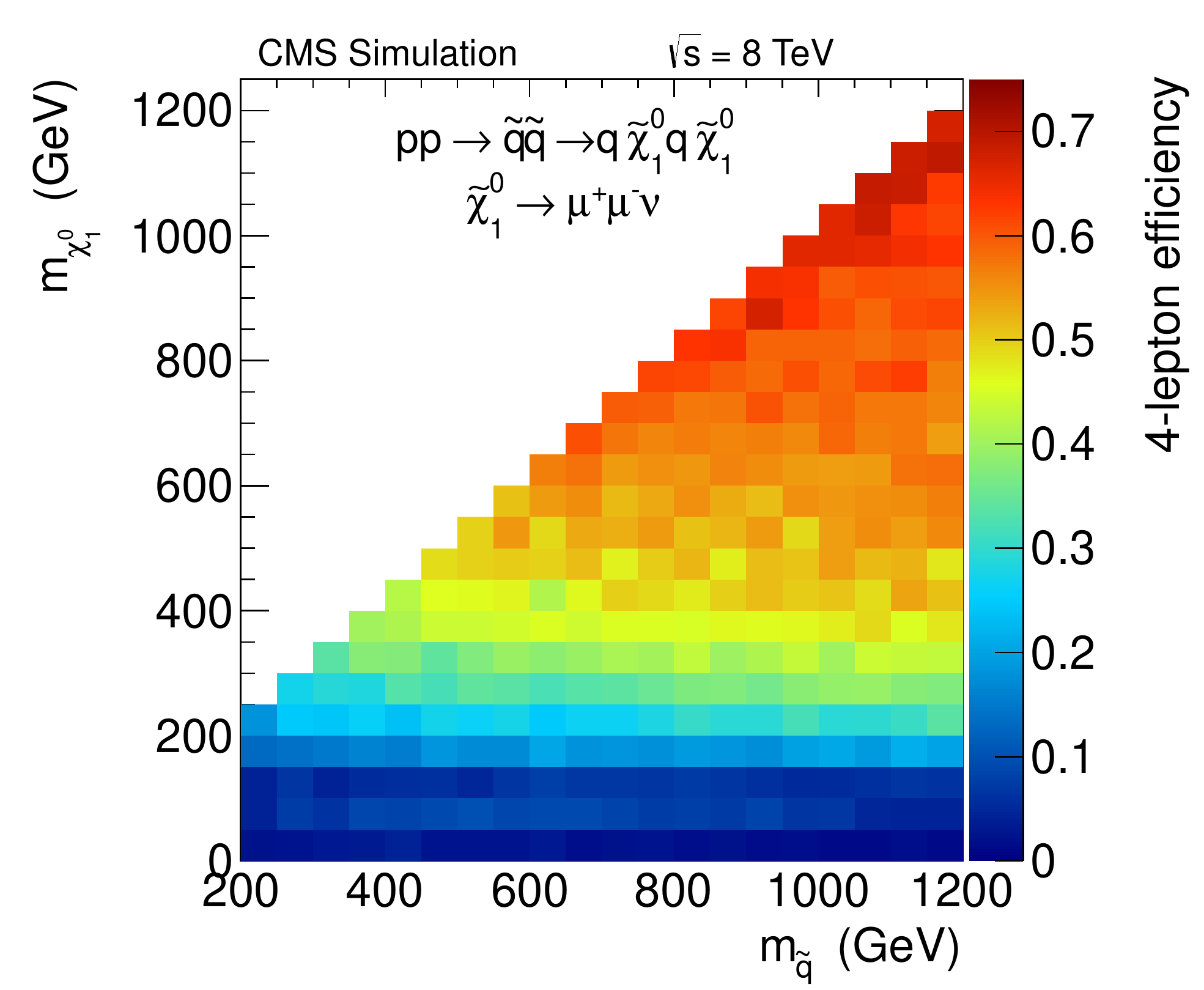}
    \includegraphics[width=0.49\textwidth]{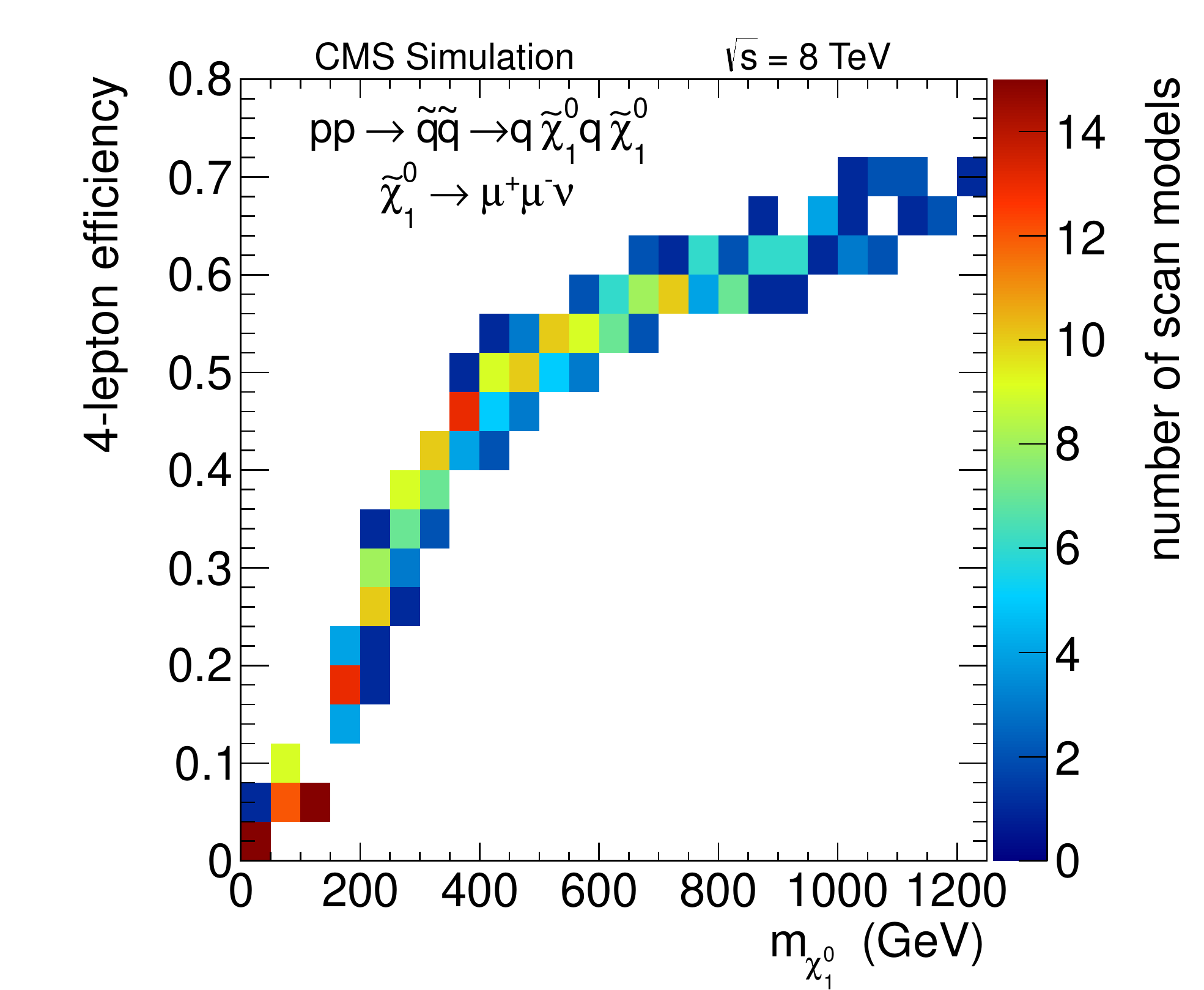}
    \caption{
      Left: efficiency for the T2+LRPV model. $\tilde{\chi}^0_1\rightarrow \mu^+\mu^-\nu$.
      Right: For every neutralino mass the efficiency value is filled corresponding to 
     the different squark masses.
    }
    \label{fig:T2LRPVeff}
  \end{center}
\end{figure}

The first approach creates the most energetic
neutralinos possible, constrained by the relevant squark and neutralino masses.
Figure~\ref{fig:T2LRPVeff} (left) presents the efficiency as a function of the T2 model parameters:
squark mass and neutralino mass. This distribution illustrates that 
the total efficiency of this analysis is mostly driven by the neutralino mass, while
the squark mass, which drives the neutralino spectrum, affects the efficiency only marginally.
To illustrate this further Fig.~\ref{fig:T2LRPVeff} (right) shows the distribution of 
the efficiency for different squark masses. 
This distribution demonstrates, that the variations even over a wide range 
of squark masses, are within $\pm$10\%.

\begin{figure}[!htdp]
  \begin{center}
    \includegraphics[width=0.49\textwidth]{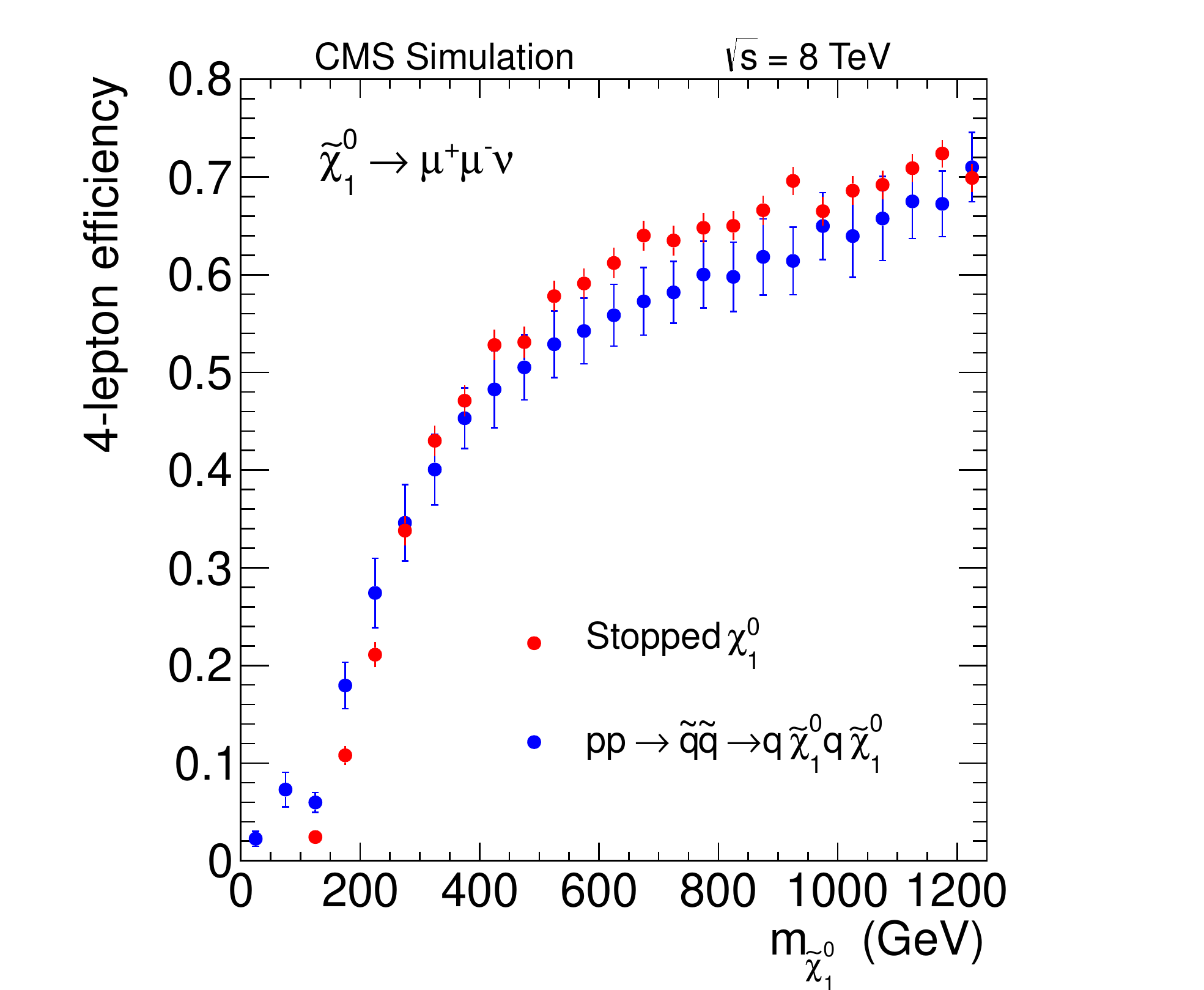}
    \hfill
    \includegraphics[width=0.49\textwidth]{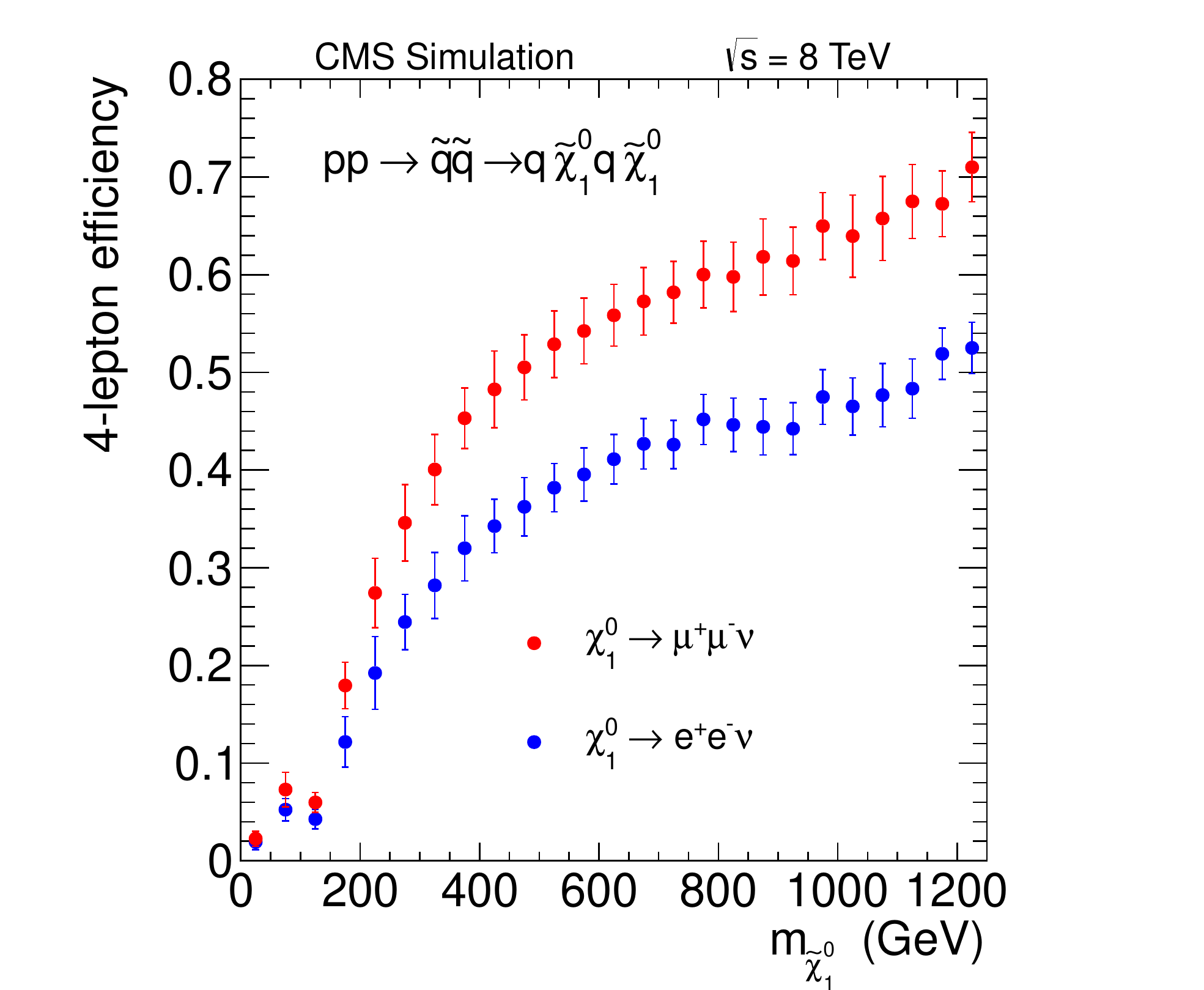}
    \caption{
      Left: efficiency of this analysis for neutralinos decaying in rest (red points), 
      overlaid with LRPV efficiency from Fig.~\ref{fig:T2LRPVeff}. 
      Right: efficiency profiles for electrons and muons. 
    }
    \label{fig:T2LRPVStopped}
  \end{center}
\end{figure}

If the LSP is produced at the end of a long cascade of decays of SUSY particles,
the LSP $p_T$ spectra will be significantly softer than for LSPs produced in two-body decays of 
the T2 scenario.
To study the effect of soft spectra we consider another extreme case: neutralino pairs produced
in rest in the detector frame. We generate the corresponding dataset by letting the neutralino 
decay into
($e^+,e^-,\nu$) or ($\mu^+,\mu^-,\nu$). Figure~\ref{fig:T2LRPVStopped} (left) shows the efficiency 
as a function of the neutralino mass overlaid with the efficiency band obtained from the T2+LRPV
model presented in Fig.~\ref{fig:T2LRPVeff} (right). It demonstrates that the difference 
between the T2+LRPV case and the stopped neutralino case is below $\pm$10\%.

\begin{figure}[!htdp]
  \begin{center}
    \includegraphics[width=0.7\textwidth]{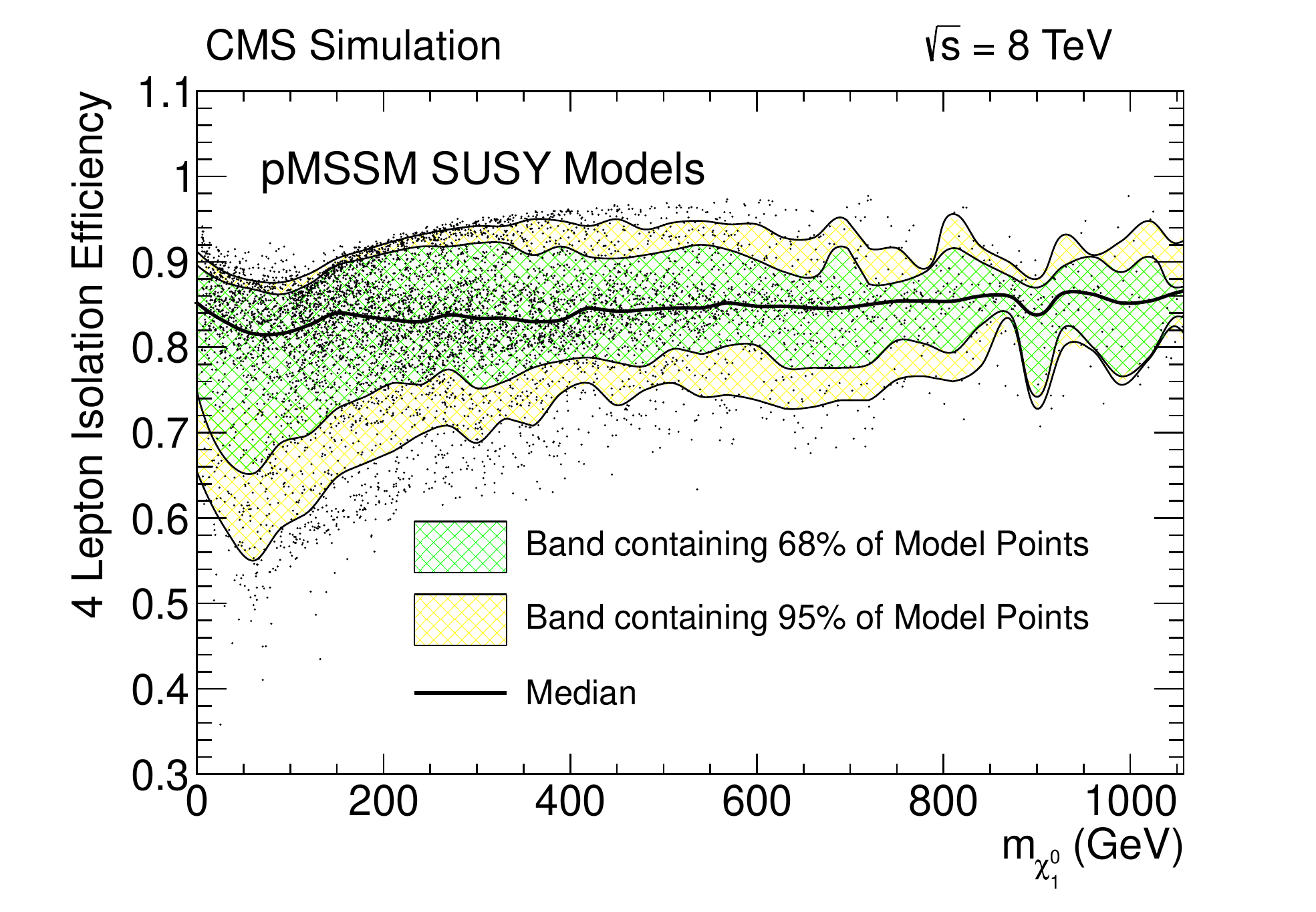}
    \caption{
      Isolation efficiency for 4 leptons for the set of pMSSM models described
    in the text, as a function of the neutralino mass in the model.
    The green and yellow bands include 68\% and 95\% of the model points in the 
    efficiency distribution respectively. 
    }
    \label{fig:pMSSM_eff_bands}
  \end{center}
\end{figure}

The isolation efficiency for isolated leptons from RPV decays depends on the occupancy of the event,
which in turn depends on the content of the underlying SUSY event. 
To study how strong the influence
of different underlying SUSY models and different SUSY production mechanisms is, 
we re-use the data samples produced in a previous CMS analysis \cite{CMS-PAS-SUS-12-030}.
These are MC samples for about 7300 different RPC phenomenological MSSM (pMSSM) \cite{Djouadi_1998di} 
 model points, each one containing 10000 events, selected to fulfill 
different pre-CMS observations. The pMSSM model is an excellent proxy for 
the full MSSM with a sufficiently small number of parameters \cite{CMS-PAS-SUS-12-030}. 
The available datasets
for this set of pMSSM models is to date the biggest sample of varying SUSY models available to us. 
To evaluate the effect of different occupancies in each event of the pMSSM, we start by extracting
the generator-level information about the neutralino. Then we generate a neutralino RPV decay 
into two leptons and a neutrino and finally
calculate the reconstruction level isolation around the direction of the obtained leptons. 
The event is accepted if the
isolation for each of the 4 charged leptons satisfies the isolation requirements for prompt leptons
used in this analysis.
Figure~\ref{fig:pMSSM_eff_bands}  presents the efficiencies for different
SUSY models as a function of the neutralino mass in each model. 
Nearly all SUSY models have a
4-lepton isolation efficiency in the range between 0.5 and 1. 
The green and yellow shaded areas in the plot contain 68\% and 95\% of the model points respectively.

We use the band $[0.5, 1]$ as a conservative estimate for possible variations
of the analysis signal efficiency due to different types of underlying SUSY models.
We use $30\%$ uncertainty when we combine this effect with other uncertainties. 

Combining all effects, we consider the T2+LRPV model efficiency in 
Fig.~\ref{fig:T2LRPVStopped} (right) to be a representative of a ``best efficiency'' scenario.
Large hadronic activity in the event can reduce the isolation efficiency. In line with the pMSSM
study, we conclude that the reduction of
the total efficiency for this search may be up to 50\%. Therefore, we consider an efficiency 
band between these two extreme 
cases to cover the 4-lepton efficiency for most the SUSY
models in this analysis.

Once an upper limit on $\sigma\times\mathcal{L}\times\varepsilon$ is extracted from the observations,
and the efficiency is evaluated, the corresponding
limit on the cross section, $\sigma^{SUSY}_{total}$, may be calculated.

\begin{figure}[!htdp]
  \begin{center}
    \includegraphics[width=0.49\textwidth]{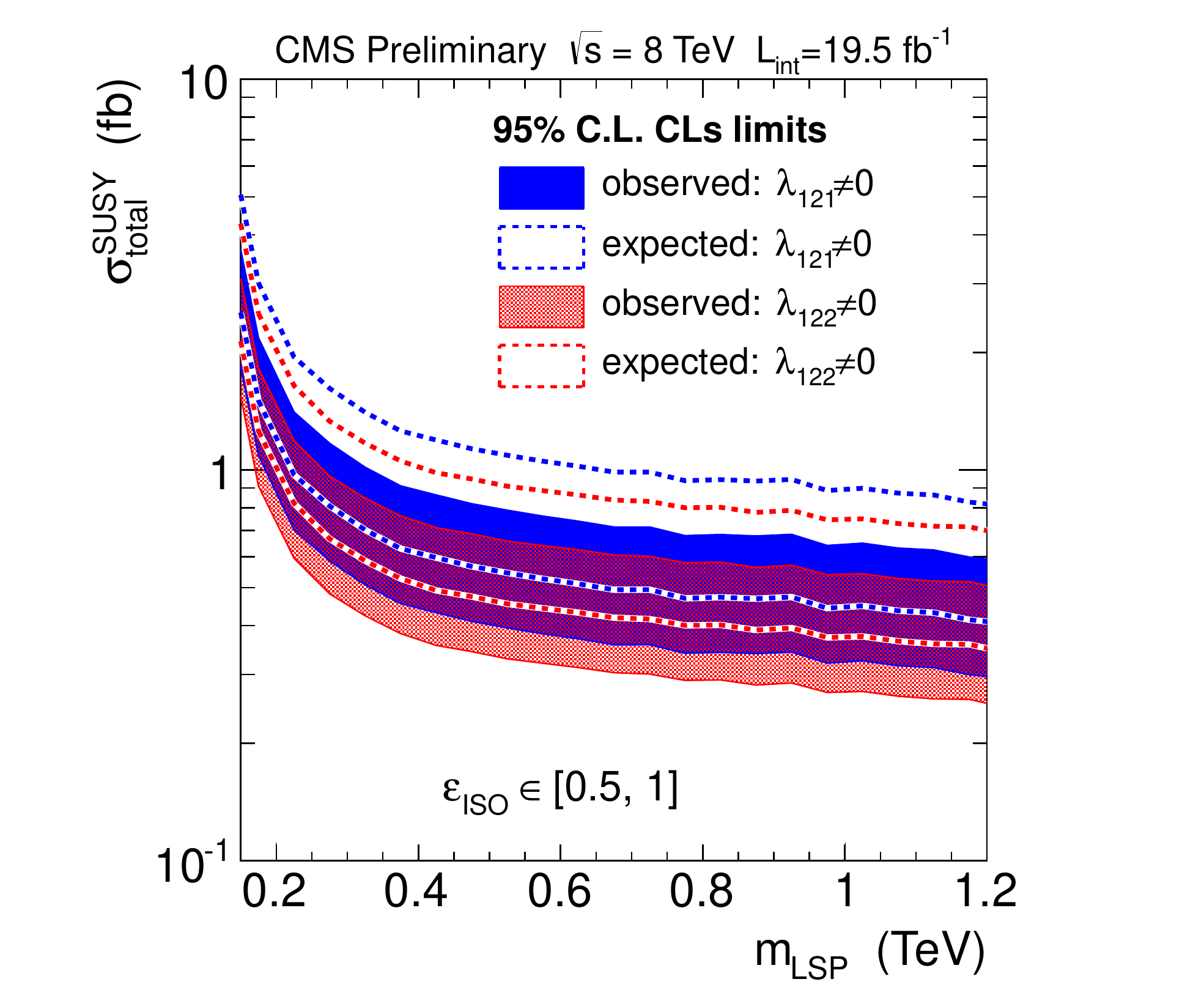}
    \hfill
    \includegraphics[width=0.49\textwidth]{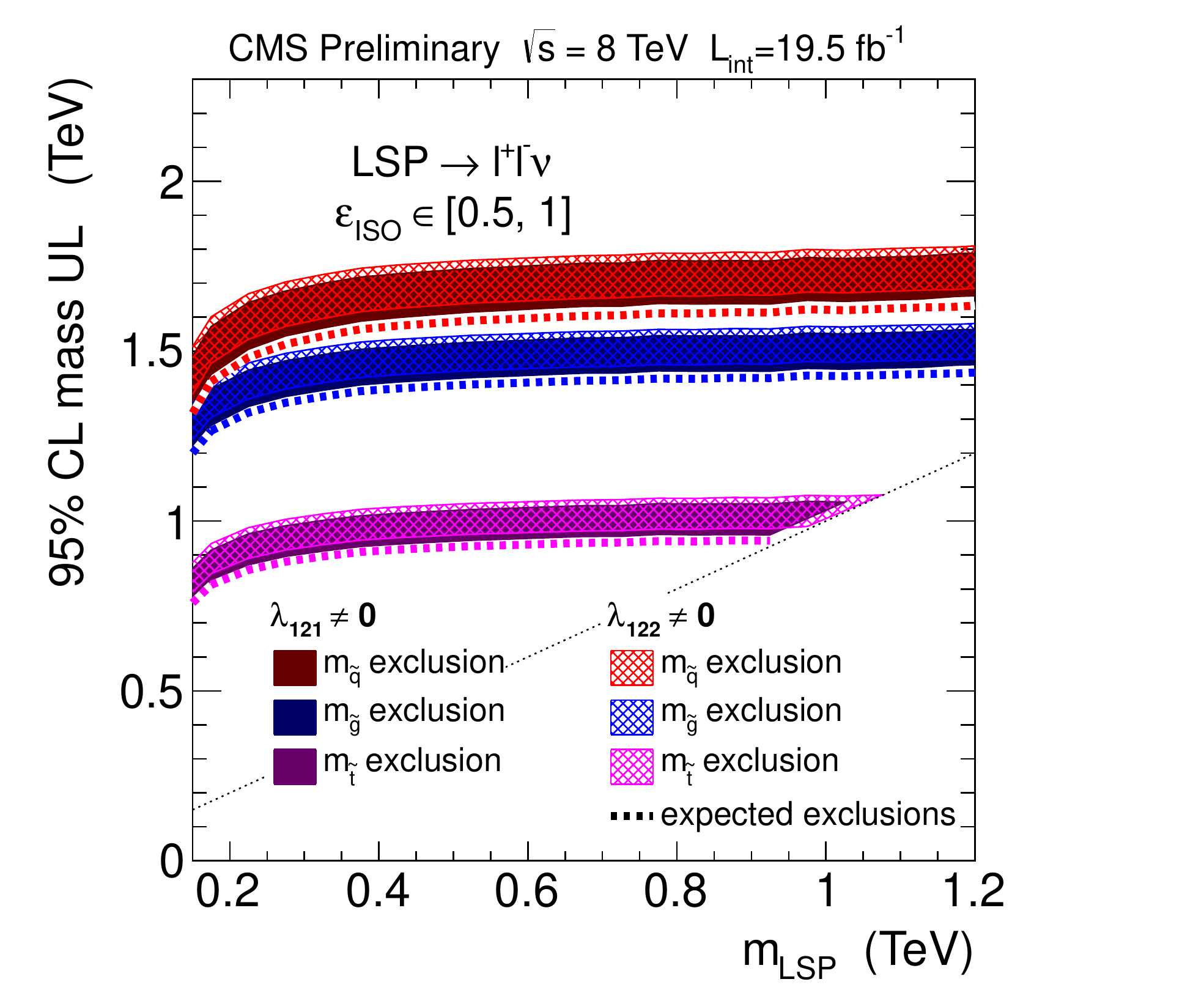}
    \caption{
      Left: 95\% C.L. upper limit on total cross sections for generic SUSY models.
   The band corresponds to the efficiency uncertainty as described in the text.
   Right: Mass exclusions for different SUSY production mechanisms. Left: for T2+LRPV models.
    Right: using a generic total RPV SUSY cross section limit in the left plot.
    A $30\%$ theoretical uncertainty for NLO+NLL calculations of SUSY production cross sections is included
    in the uncertainty band.
    }
    \label{fig:massBands}
  \end{center}
\end{figure}

The experimental observations together with
the pMSSM based efficiency estimation as described above 
drive the exclusion for the cross section of total RPV SUSY production,
which is presented in Fig.~\ref{fig:massBands} (left). The bands correspond to 
the 4 lepton isolation variations between 50\% and 100\%.
Note that this is a very generic result as this band covers 
RPV models with a wide range of underlying RPC SUSY models.

To further convert the cross section limit into a mass exclusion we consider several SUSY 
production mechanisms:
gluino pair production, squark pair production, and stop-quark pair production.
The cross sections for these processes as functions of the corresponding masses 
are NLO+NLL calculation results of the corresponding decoupled scenarios \cite{Kramer_2012bx}. 
The theoretical uncertainties on the NLO+NLL SUSY production cross section calculations for masses $\sim$1\TeV
are about $30\%$, and are accounted for in the result. 

Using these total cross sections as a function of the mass of the corresponding SUSY particle,
we convert the cross section limit bands in Fig.~\ref{fig:massBands} (left) into mass exclusion
bands as a function of the LSP mass. This result is presented in Fig.~\ref{fig:massBands} (right).

\section{Conclusions}
CMS developed a comprehensive program for RPV SUSY searches. In this contribution we present the most
recent results on this topic. For all presented searches observations are consistent
with expectations from the Standard Model, thus the corresponding limits on presence of new physics are set.
We also present a new approach of generalizing physics interpretations of experimental observations.
Sampling of a big set of pMSSM models allows to check a model dependency for obtained results,
thus making more general conclusions possible. 

%
%
%

\end{document}